\documentclass[floatfix,aps,twocolumn,preprintnumbers,amsmath,amssymb,nofootinbib,superscriptaddress]{revtex4}

\usepackage{graphicx,color}
\usepackage{amsmath,amssymb,bm}
\usepackage{natbib} 

\def\1{\'{\i}}

\def\XXint#1#2#3{{\setbox0=\hbox{$#1{#2#3}{\int}$}
     \vcenter{\hbox{$#2#3$}}\kern-.5\wd0}}

\newcommand{\nk}{{\bf k}}
\newcommand{\np}{{\bf p}}
\newcommand{\nq}{{\bf q}}

\def\XXint#1#2#3{{\setbox0=\hbox{$#1{#2#3}{\int}$}
     \vcenter{\hbox{$#2#3$}}\kern-.5\wd0}}

\def\1{\'{\i}}

\begin{document}


\title{
Quasielastic charged-current neutrino scattering 
in the scaling model  with relativistic effective mass
}


\author{I. Ruiz Simo}\email{ruizsig@ugr.es} \affiliation{Departamento de
  F\'{\i}sica At\'omica, Molecular y Nuclear \\ and Instituto Carlos I
  de F{\'\i}sica Te\'orica y Computacional \\ Universidad de Granada,
  E-18071 Granada, Spain.}

\author{V.L. Martinez-Consentino}
 \affiliation{Departamento de
  F\'{\i}sica At\'omica, Molecular y Nuclear \\ and Instituto Carlos I
  de F{\'\i}sica Te\'orica y Computacional \\ Universidad de Granada,
  E-18071 Granada, Spain.}

\author{J.E. Amaro}\email{amaro@ugr.es} \affiliation{Departamento de
  F\'{\i}sica At\'omica, Molecular y Nuclear \\ and Instituto Carlos I
  de F{\'\i}sica Te\'orica y Computacional \\ Universidad de Granada,
  E-18071 Granada, Spain.}

\author{E. Ruiz
  Arriola}\email{earriola@ugr.es} \affiliation{Departamento de
  F\'{\i}sica At\'omica, Molecular y Nuclear \\ and Instituto Carlos I
  de F{\'\i}sica Te\'orica y Computacional \\ Universidad de Granada,
  E-18071 Granada, Spain.}


\begin{abstract}
We use a recent scaling analysis of the quasielastic electron
scattering data from $^{12}$C to predict the quasielastic
charge-changing neutrino scattering cross sections within an
uncertainty band. We use a scaling function extracted from
a selection of the $(e,e')$ cross section data, and an
effective nucleon mass inspired by the relativistic mean-field model
of nuclear matter.  The corresponding super-scaling analysis with
relativistic effective mass (SuSAM*) describes a
large amount of the electron data lying inside a phenomenological
quasielastic band.  The effective mass incorporates the enhancement of
the transverse current produced by the relativistic mean field. 
The scaling function incorporates
nuclear effects beyond the impulse approximation, in particular
meson-exchange currents and short range correlations producing tails
in the scaling function. Besides its simplicity, this model describes
the neutrino data as reasonably well as other more sophisticated
nuclear models.

\end{abstract}


\maketitle

\section{Introduction}

The analysis of modern accelerator-based neutrino oscillation
experiments requires a precise knowledge of the intermediate-energy
neutrino-nucleus scattering cross section \cite{Mos16,Kat17}.  The
inclusive cross section involves contributions from different
channels, which can be grouped into quasielastic (QE) one-nucleon
emission, multi-nucleon (2p-2h, \ldots) emission, pion production and
other inelastic processes. In particular, QE interactions are key
processes for these experiments, but the nuclear models for these
neutrino and antineutrino cross sections have large uncertainties
\cite{Pat18}.  There are processes which appear as QE-like in the
neutrino detectors, produced by final-state interactions, direct
multi-nucleon emission, and others which are not under total control
from the theoretical point of view, and require relativistic modeling
of complex hadronic final states in the continuum.  They therefore
limit the reach of current and future oscillation experiments such as
T2K \cite{Abe11,Abe18a,Abe18b}, NOvA \cite{Ada16} or DUNE
\cite{Acc15}.  Recent attempts to reduce the model uncertainties have
been made by measuring the proton multiplicity of the final states in
the T2K \cite{Abe18b} and ArgoNEUT \cite{Pal16,Acc14} experiments and
also the measurement of neutron multiplicity in ANNIE \cite{Bac17} is
planned.

Within this state of afairs the acquaintance of the
inclusive $(e,e')$ cross sections of nuclei becomes a valuable
starting point; its prior description should be a very convenient
requisite for the neutrino-nucleus interaction models. In fact, the
isovector component of the electromagnetic nuclear responses can be
related to the vector-vector (VV) component of the weak
charged-current responses contributing to the neutrino cross sections
for the same intermediate energies. The contribution of the axial
current might, in principle, be inferred by starting from any
available model capable of describing electron scattering.  While
results from different groups including effects beyond the impulse
approximation have allowed to explain the recent neutrino and
antineutrino data \cite{Mar09,Nie11,Gal16,Meg16}, systematic
differences still persist between these theoretical predictions. It is
therefore reasonable to suspect that these differences might be
attributed to systematic differences in the description of the
$(e,e')$ data by the same models.

 The goal of this paper is to provide predictions for neutrino cross
 sections with their systematic error inherited directly from the
 available $(e,e')$ data in the superscaling model with relativistic
 effective mass (SuSAM*) \cite{Ama15,Ama17}.  The scaling approaches
 \cite{Alb88,Day90,Don99} are an alternative to more sophisticated
 microscopical models for predicting the neutrino QE cross section.
 They use a phenomenological scaling functions extracted from the
 $(e,e')$ data, which encode many effects and assume that the same
 scaling functions can be used to compute the neutrino cross section
 \cite{Ama05a,Ama05b} by just replacing the electromagnetic currents
 with the vector and axial ones.

The SuSAM* approach used in this paper is an interesting new
alternative to the more traditional superscaling analysis approach
(SuSA) of refs. \cite{Meg16a,Meg18}, using different assumptions and
definitions for the scaling functions and variables.

SuSAM* is based on the relativistic mean field or Walecka model of
nuclear matter \cite{Ser86}, containing basic theoretical and
phenomenological ingredients such as relativity, gauge invariance and
dynamical enhancement of lower Dirac components of the nucleon in the
medium due to the scalar and vector potentials. These are known to be
good enough to describe the electromagnetic nuclear responses in the
QE peak \cite{Ros80}.

Recently, we have applied SuSAM* to extract a phenomenological scaling
function directly from the cross section $(e,e')$ data in the QE
region within an uncertainty band \cite{Ama17b}.  In
\cite{Ama17b,Con18} we have shown that the extracted scaling function
$f^*(\psi^*)$ is an universal function valid for all the nuclei,
provided that a relativistic effective mass $M^*$ and Fermi momentum
$k_F$ are fitted to the data for each nucleus. These two parameters,
$k_F$ and $M^*$, have been determined in \cite{Ama17b} from the $(e,e')$
database representing a first direct extraction of the Fermi-momentum
dependence of relativistic effective mass below saturation from finite
nuclei. We find that a subset of a third of the about 20000 existing data
approximately scales to an universal superscaling function.  The
resulting scaling function and its uncertainty band have been
parameterized and can thus be easily and directly applied to {\it
  predict} the neutrino QE cross sections within a corresponding
uncertainty.

Before proceeding further a qualitying remark regarding interpretation
of the present work is in order. From our point of view the SuSAM*
band can be understood as the uncertainty in the theoretical
description of the QE data due to processes violating the scaling
model assumptions. Namely, those interactions which are beyond the
impulse approximation and break the factorization of the cross
section, but that imply small corrections to the center value and
therefore can be regarded as QE-like interactions.  The
uncertainties obtained in this work for the neutrino cross sections
can then be considered most likely as an upper limit to the systematic
error expected from nuclear modeling of the QE processes, because all
the models aiming to describe the $(e,e')$ data should lie inside the
phenomenological uncertainty bands for the pertinent kinematics.

The structure of this work is as follows. In Sect. II we describe the
theoretical formalism of the SuSAM* model and the parameters of the
phenomenological superscaling function $f^*(\psi^*)$. In Sect. III we give
our results for the neutrino and antineutrino cross sections and
theoretical uncertainties and compare with most of the available data
sets. In sect. IV we give our summary and conclusions.
In the appendices we show some technical details on the calculation of
selected differential cross sections.

\section{Formalism of quasielastic neutrino scattering}

\subsection{Cross section and responses}

In this paper we are interested in the charged-current quasielastic
(CCQE) reactions in nuclei induced by neutrinos.  In particular we
compute the $(\nu_\mu,\mu^-)$ cross section.  The total energies of
the incident neutrino and detected muon are $\epsilon=E_\nu$,
$\epsilon'=m_\mu+T_\mu$, and their momenta are $\nk,\nk'$.  The
four-momentum transfer is $k^\mu-k'{}^\mu=(\omega,\nq)$, with
$Q^2=q^2-\omega^2 > 0$.

If
the lepton scattering angle is $\theta_\mu$, the double-differential cross
section can be written as \cite{Ama05a,Ama05b}
\begin{eqnarray}
\frac{d^2\sigma}{dT_\mu d\cos\theta_\mu}
&=&
\sigma_0
\left\{
V_{CC} R_{CC}+
2{V}_{CL} R_{CL}
\right.
 \nonumber\\
&&
\left.
+{V}_{LL} R_{LL}+
{V}_{T} R_{T}
\pm
2{V}_{T'} R_{T'}
\right\} \, , 
\end{eqnarray}
where we have defined the cross section 
\begin{equation}
\sigma_0=
\frac{G^2\cos^2\theta_c}{4\pi}
\frac{k'}{\epsilon}v_0. 
\end{equation}
Here $G=1.166\times 10^{-11}\quad\rm MeV^{-2} \sim 10^{-5}/ m_p^2$ is
the Fermi constant, $\theta_c$ is the Cabibbo angle,
$\cos\theta_c=0.975$, and the kinematic factor $v_0=
(\epsilon+\epsilon')^2-q^2$.
The nuclear structure is implicitly written
as  a linear combination of five nuclear response functions,  $R_K(q,\omega)$, 
where the fifth response function $R_{T'}$ is added
 (+)  for neutrinos and subtracted ($-$) for antineutrinos).
The $V_K$ coefficients depend only on the lepton kinematics 
and are independent on the details of the nuclear target. They are defined by
\begin{eqnarray}
{V}_{CC}
&=&
1-\delta^2\frac{Q^2}{v_0}
\label{vcc}\\
{V}_{CL}
&=&
\frac{\omega}{q}+\frac{\delta^2}{\rho'}
\frac{Q^2}{v_0}
\\
{V}_{LL}
&=&
\frac{\omega^2}{q^2}+
\left(1+\frac{2\omega}{q\rho'}+\rho\delta^2\right)\delta^2
\frac{Q^2}{v_0}
\\
{V}_{T}
&=&
\frac{Q^2}{v_0}
+\frac{\rho}{2}-
\frac{\delta^2}{\rho'}
\left(\frac{\omega}{q}+\frac12\rho\rho'\delta^2\right)
\frac{Q^2}{v_0}
\\
{V}_{T'}
&=&
\frac{1}{\rho'}
\left(1-\frac{\omega\rho'}{q}\delta^2\right)
\frac{Q^2}{v_0}.
\label{vtp}
\end{eqnarray}
Here we have defined the dimensionless factors $\delta =
m_\mu/\sqrt{Q^2}$, proportional to the muon mass $m_\mu$, $\rho =
Q^2/q^2$, and $\rho' = q/(\epsilon+\epsilon')$.

We evaluate the five nuclear response functions $R_K$, $K=CC, CL, LL,
T, T'$ ($C$=Coulomb, $L$=longitudinal, $T$=transverse)
using a coordinate system with the $z$-axis pointing along $\nq$ and the
$x$-axis along the transverse component of the incident neutrino. 
The nuclear response functions in this frame are given by the following components of the hadronic tensor:
\begin{eqnarray}
R_{CC } & = & W^{00 }  \\
R_{CL } & = & -\frac12 (W^{03 } + W^{30 } ) \\
R_{LL } & = & W^{33 }  \\
R_{T } & = & W^{11 } + W^{22 } \\
R_{T' } & = & -\frac{i}{2}(W^{12 } - W^{21 }).
\end{eqnarray}

\subsection{Nuclear matter responses in the RMF }

The starting point in this work is the relativistic mean field (RMF)
theory of nuclear matter \cite{Ser86}, and its reasonable 
description of the electromagnetic nuclear
response in the quasielastic region \cite{Ros80}.
This model \cite{Ser86} describes the nuclear interaction in terms of 
vector and scalar potentials whose effect is encoded into a
relativistic effective mass $m^*_N$ of the nucleon in the medium.

In this model the
hadronic tensor for one particle-one hole (1p-1h) excitations
with momentum $\nq$ and energy $\omega$ 
can be written as
\begin{eqnarray}
W^{\mu\nu}(q,\omega) 
&=& \frac{V}{(2\pi)^3} \int d^3p \delta(E'-E-\omega)
\frac{(m_N^*)^2}{EE'} 
\nonumber\\
&\times& 
2w^{\mu\nu}_{s.n.}(\np',\np)
\theta(k_F-p)\theta(p'-k_F)
\label{hadronictensor}
\end{eqnarray}
where 
$E=\sqrt{\np^2+m_N^*{}^2}$ is the initial nucleon energy in the mean
field. The final momentum of the nucleon is $\np'=\np+\nq$ and its
energy is $E'=\sqrt{\np'{}^2+m_N^*{}^2}$. Note that the initial and final
nucleons have the same effective mass $m_N^*$.  
The volume $V=3\pi^2 {\cal N}/k_F^3$
of the system is related to the Fermi momentum $k_F$ and proportional to
the number ${\cal N}=N (Z) $ of neutrons (protons)  participating in the
process for CC neutrino (antineutrino) scattering. 
Finally the single-nucleon tensor is written in terms of the CC current
\begin{equation}
w^{\mu\nu}_{s.n.}(\np',\np)=\frac12\sum_{ss'}J^{\mu *}(\np',\np)J^{\nu}(\np',\np)
\label{singlenucleontensor}
\end{equation}
where $J^{\mu *}$ is the weak current matrix element between
 positive energy Dirac spinors with mass $m_N^*$ and
normalized to $\overline{u}u = 1$. 
This single nucleon current is the sum of vector and axial-vector terms
$J^\mu= V^\mu- A^\mu $
where the vector current is
\begin{equation}
V^\mu_{s's}=
\overline{u}_{s'}(\np')
\left[ 
2F_1^V\gamma^\mu 
+2F_2^Vi\sigma^{\mu\nu}\frac{Q_\nu}{2m_N}
\right]u_{s}(\np)
\label{vector}
\end{equation}
where $F_i^V=(F_i^P-F_i^N)/2$ are the isovector form factors of the nucleon.
The axial current is
\begin{equation}\label{eq:axial_current_def}
A^\mu_{s's}=
\overline{u}_{s'}(\np')
\left[ 
G_A\gamma^\mu\gamma_5 
+G_P\frac{Q^\mu}{2m_N}\gamma_5
\right]u_{s}(\np)
\end{equation}
Note that the free nucleon mass enters in the current operator,
which is not modified in the medium.
However
the initial and final spinors 
$u_{s}(\np)$, and $u_{s'}(\np')$, 
correspond to  nucleons with relativistic effective mass
$m^*_N$. This modifies the values of the above matrix elements in the
nuclear medium with respect to the free values. 
Thus the vector current operator used in this work, Eq. (\ref{vector})  corresponds
to the CC2 prescription for the off-shell extrapolation 
of the electromagnetic current operator 
\cite{DeF83}.

\begin{figure*}
\centering
\includegraphics[width=\textwidth]{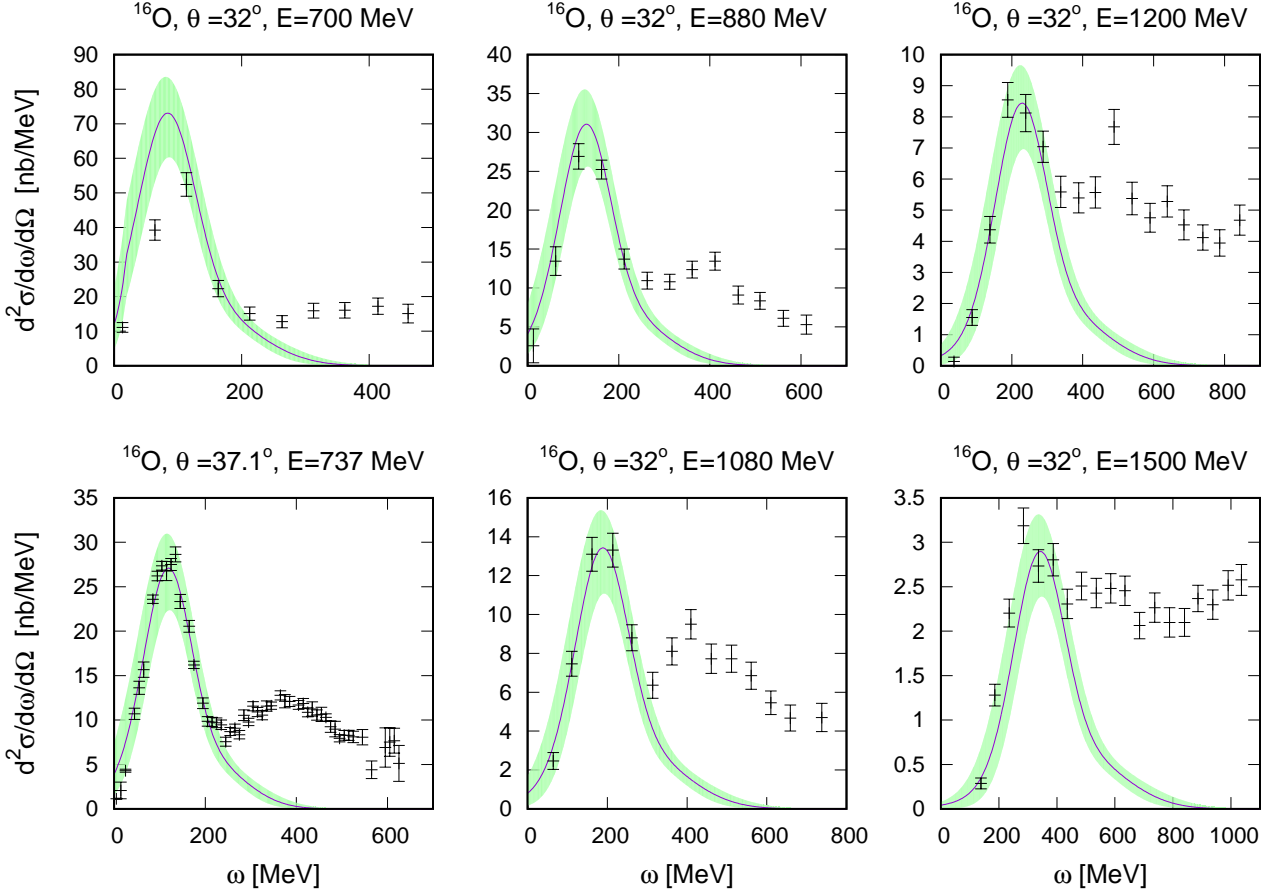}
\caption{Comparison of inclusive $^{16}$O $(e,e')$ cross section and
  predictions of the SuSAM* model. The solid lines have been obtained
  with the central parameterization of the scaling function,l while the
  green band represents the theoretical uncertainty. 
Data are from \cite{Ang96,Con87}.}\label{electron}
\end{figure*}

The present RMF theory of nuclear matter treats exactly relativity,
gauge invariance and translational invariance. It differs with respect
to the well known RFG in that the nucleon mass is replaced by the
effective mass both in the spinors and in the energy-momentum
relation. As a consequence, equations similar to the RFG responses are
obtained by replacing $m_N$ by $m_N^*$ and re-scaling some of the
form factors.

The resulting nuclear response function $R_K$ 
is proportional to a single-nucleon response function $U_K$ times the
 scaling function $f^*(\psi^*)$
\begin{equation} \label{rfg}
R_K = \frac{{\cal N} \xi^*_F}{m^*_N \eta_F^{*3} \kappa^*}  U_K  f^*(\psi^*)
\end{equation}
where ${\cal N}$ is the number of neutrons/protons for
neutrino/antineutrino scattering, $\eta^*_F=k_F/m^*_N$, and
$\xi^*_F=\sqrt{1+\eta_F^{*2}}-1$.  This factorization of the scaling
function inspires the scaling models of \cite{Alb88,Day90,Don99}, by
using a phenomenological scaling function instead of the well-known
scaling function of the Fermi gas
\begin{equation}
f_{\rm RFG}^*(\psi^*)=\frac34 (1-\psi^{*2})\theta(1-\psi^{*2}) \, , 
\end{equation}
where $\theta $ is the step function and $\psi^*$ is the scaling
variable. In this work we use a phenomenological
scaling function extracted in \cite{Ama17} from the $(e,e')$ 
quasielastic data {\bf of ${}^{12}$C.}

In contrast to traditional approaches where the lepton interacts with a
 free nucleon, here we use a modified scaling variable incorporating the effective mass
\begin{equation}
\psi^{*2} = \frac{1}{\xi^*_F} {\rm max}
\left\{
\kappa^*\sqrt{1+\frac{1}{\tau^*}}-\lambda^* -1 , \xi^*_F-2\lambda^*
\right\}
\end{equation}
where $\lambda^*=\omega/(2m^*_N)$, $\kappa^*=q/(2m^*_N)$, and 
$\tau^*=\kappa^{*2}-\lambda^{*2}$.

The single-nucleon responses $U_K$ are obtained analytically
by performing the traces in Eq. (\ref{singlenucleontensor}) and the
integration in Eq. (\ref{hadronictensor}).

The $K=CC$ response is the sum of vector and axial pieces. The vector
part implements the conservation of the vector current (CVC), i.e.  it
vanishes by contracting with $q^\mu$. The axial part can be written
as the sum of conserved (c.)  plus non conserved (n.c.) parts. Then
\begin{eqnarray} \label{ucc}
U_{CC} &=& U_{CC}^{V}+
\left(U_{CC}^{A}\right)_{\rm c.}
+\left(U_{CC}^{A}\right)_{\rm n.c.}
\end{eqnarray}
For the vector CC response we have
\begin{eqnarray}
U_{CC}^V &=&
\frac{\kappa^{*2}}{\tau^*}
\left[ (2G_E^{*V})^2+\frac{(2G_E^{*V})^2+\tau^* (2G_M^{*V})^2}{1+\tau^*}\Delta
\right]\ ,
\label{uccv}
\end{eqnarray}
where $G_E^{*V}$ and $G_M^{*V}$ are the new isovector electric and magnetic
nucleon form factors that get modified in the medium 
through the effective mass:
\begin{eqnarray}
G_E^{*V}  &=&  F_1^V-\tau^* \frac{m^*_N}{m_N} F_2^V \\
G_M^{*V}  &=& F_1^V+\frac{m_N^*}{m_N} F_2^V.  \label{GM}
\end{eqnarray}
For the free Dirac and Pauli form
factors, $F_1^V$ and $F_2^V$, we use the Galster parameterization.

The definition of the quantity $\Delta$ in Eq. (\ref{uccv}) is
\begin{equation}
\Delta= \frac{\tau^*}{\kappa^{* 2}}\xi^*_F(1-\psi^{*2})
\left[\kappa^*\sqrt{1+\frac{1}{\tau^*}}+\frac{\xi^*_F}{3}(1-\psi^{*2})\right].
\end{equation}
The axial-vector CC responses are 
\begin{eqnarray}
\left(U_{CC}^A\right)_{\rm c.}
&=&
\frac{\kappa^{*2}}{\tau^*}G_A^2\Delta
\\
\left(U_{CC}^A\right)_{\rm n.c.}
&=&
\frac{\lambda^{*2}}{\tau^*}(G_A - \tau^* G_P^* )^2.
\end{eqnarray}
where $G_A$ is the nucleon axial-vector form factor and $G_P^*$ is the
new pseudo-scalar axial form factor, also modified in the medium.
From partial conservation of the axial current (PCAC), the new and
re-scaled pseudo-scalar form factor are now
\begin{equation}
G^*_P =  \frac{4m_Nm_N^*}{m_\pi^2+Q^2}G_A.
\end{equation}
Note that the axial form factor $G_A$ is not modified in the medium
because in its definition in the axial current,
Eq.~(\ref{eq:axial_current_def}), the nucleon mass does not appear
explicitly.

Using current conservation  we have  for $K=CL,LL$
\begin{eqnarray}
U_{CL} &=& -\frac{\lambda^*}{\kappa^*}
\left[U_{CC}^V+\left(U_{CC}^A\right)_{\rm c.}\right]
+\left(U_{CL}^A\right)_{\rm n.c.}
\label{ucl} \\
U_{LL} &=& 
\frac{\lambda^{*2}}{\kappa^{*2}}
\left[ U_{CC}^V+\left(U_{CC}^A\right)_{\rm c.} \right]
+\left(U_{LL}^A\right)_{\rm n.c.}\ ,
\label{ull}
\end{eqnarray}
The n.c. parts are
\begin{eqnarray}
\left(U_{CL}^A\right)_{\rm n.c.}
&=& -\frac{\lambda^*\kappa^*}{\tau^*}(G_A - \tau^* G_P^* )^2
\\
\left(U_{LL}^A\right)_{\rm n.c.}
&=& \frac{\kappa^{*2}}{\tau^*}(G_A - \tau^* G^*_P )^2\ .
\end{eqnarray}
Finally, the transverse responses are given by
\begin{eqnarray}
U_T &=& U_T^V+U_T^A  \label{ut}
\\
U_T^V &=&  2\tau^*(2G_M^{*V})^2+\frac{(2G_E^{*V})^2+\tau^* 
(2G_M^{*V})^2}{1+\tau^*}\Delta \label{utv}
\\
U_T^A &=& G_A^2 \left[2(1+\tau^*)+ \Delta\right] \label{uta}
\\
U_{T'} &=& 2G_A(2G_M^{*V}) \sqrt{\tau^*(1+\tau^*)}[1+\tilde{\Delta}]
\label{utp}
\end{eqnarray}
with
\begin{equation}
\tilde{\Delta}=
\sqrt{\frac{\tau^*}{1+\tau^*}}\frac{\xi^*_F(1-\psi^{*2})}{2\kappa^*}\ .
\end{equation}

\subsection{SuSAM* scaling function}

One of the inputs of our model is the phenomenological scaling
function $f^*(\psi^*)$.  This has been determined from a scaling
analysis of the quasielastic electron scattering data \cite{Ama17,
  Ama15} based on the RMF formulae of the previous section, with an
effective mass for the nucleon, in contrast to all previous
investigations made with $M^*=1$ \cite{Day90,Don99,Ama05a}.  This
analysis allows to select a subset of ``quasielastic'' data which
merge into a thick band that can be separated from the rest of
data. This subset turns out to be a large fraction (about a third) of
the total 20000 data and approximately scales to an universal
superscaling function with uncertainties.  The central value of the
phenomenological quasielastic band has been parameterized as a sum of
two Gaussian functions
\begin{equation}\label{fitpar}
f^*(\psi^*)  = 
a_3 e^{-(\psi^*-a_1)^2/(2 a_2^2)}+b_3 e^{-(\psi^*-b_1)^2/(2b_2^2)}.
\end{equation}
The coefficients encoding the band and their ranges are provided in
table \ref{table_coeff}.  The lower and upper limits of the
phenomenological band have been parameterized as sum of two Gaussians
as well, with coefficients $a_i^{min/max}$, $b_i^{min/max}$, given in
table \ref{table_coeff} too.  Note that the new scaling function
$f^*(\psi^*)$ is different to the phenomenological scaling function
used in the SuSA formalism \cite{Mai02}.

\begin{table}
\begin{tabular}{crrrrrr}\hline
   & $a_1$ & $a_2$ & $a_3$ & $b_1$ & $b_2$ & $b_3$ \\ \hline
central 
&-0.0465
& 0.469
& 0.633
& 0.707
& 1.073
& 0.202
\nonumber\\
min
&-0.0270
& 0.442
& 0.598
& 0.967
& 0.705
& 0.149
\nonumber\\
max
& -0.0779
& 0.561
& 0.760
& 0.965
& 1.279
& 0.200
\\ \hline
\end{tabular}
\caption{Parameters of our fit of the phenomenological 
scaling function central value, $f^*(\psi^*)$, 
and of the lower and upper boundaries (min and max, respectively).} 
\label{table_coeff}
\end{table}

The present SuSAM* model provides a landmark representation of the
intermediate energy quasielastic data, in the sense that any model
aiming to describe these quasielastic data should lie inside the
uncertainty band.  It gives a fair and simple description of the
selected data band.  Besides the scaling function $f^*(\psi^*)$, it
only includes two parameters: the effective mass and the Fermi
momentum.

In the present work we will apply this model to neutrino and
antineutrino quasielastic scattering from $^{12}$C and $^{16}$O,
assuming the same uncertainty band in the scaling function as
determined in $(e,e')$. In ref. \cite{Ama17} we showed the results of
our model for the doubly differential $^{12}$C$(e,e')$ cross section,
with a fair global description of data. In Fig. \ref{electron} we
compare our model with the $^{16}$O data. We remark that only six
kinematics are available for this nucleus. 
We use the values $M^*=0.8$ and $k_F=230$ MeV/c, taken from the
 recent $A$-dependent super-scaling analysis of \cite{Ama17b},
 while the scaling function band is the same as that of $^{12}$C 
determined in \cite{Ama17} and given in Table \ref{table_coeff}.

As it can be observed most of the data around the quasielastic region
are well described.  Note that our model does not include the pion
emission or inelastic channels and therefore the higher energy data
lie, as expected, outside our uncertainty band.  The quality of our
description is similar to the one of the SuSAv2 model
\cite{Meg16a,Meg18}.  In addition, we provide an estimation of the
theoretical error in the cross section, that will be translated to
neutrino cross section error bands in the results of the next section.

\section{NUMERICAL PREDICTIONS}

\begin{figure*}[th]
\centering
\includegraphics[width=\textwidth]{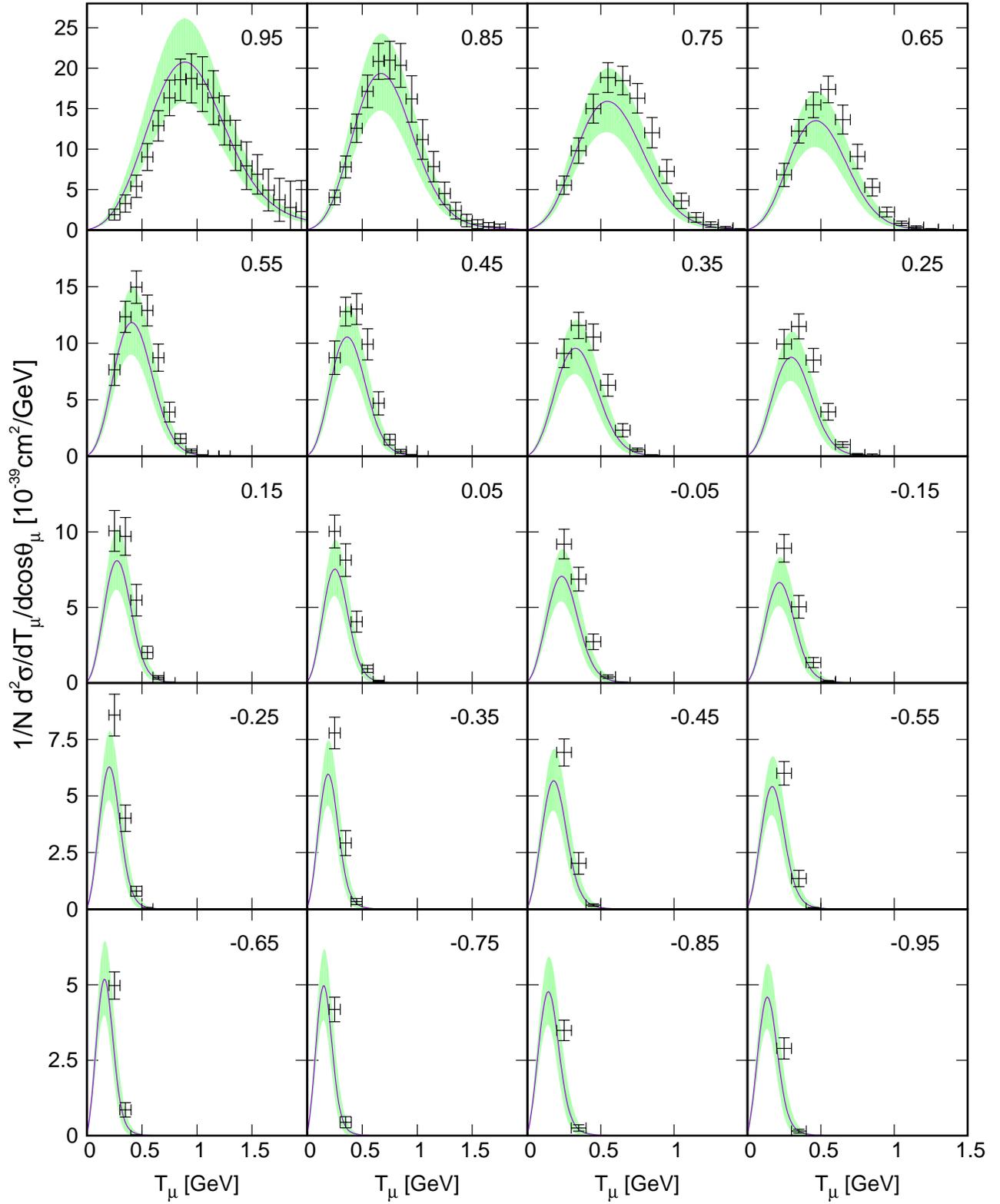}
\caption{ 
(Color online) Flux-integrated double-differential cross
  section per target neutron for the CCQE $(\nu_\mu,\mu^-)$ reaction
  on $^{12}$C in the SuSAM* model. 
Each panel is labeled by
  the mean value of $\cos\theta_\mu$ in each experimental bin. 
The experimental data are from \cite{Agu10}
}\label{nu-miniboone}
\end{figure*}

\begin{figure*}[thp]
\centering
\includegraphics[width=\textwidth]{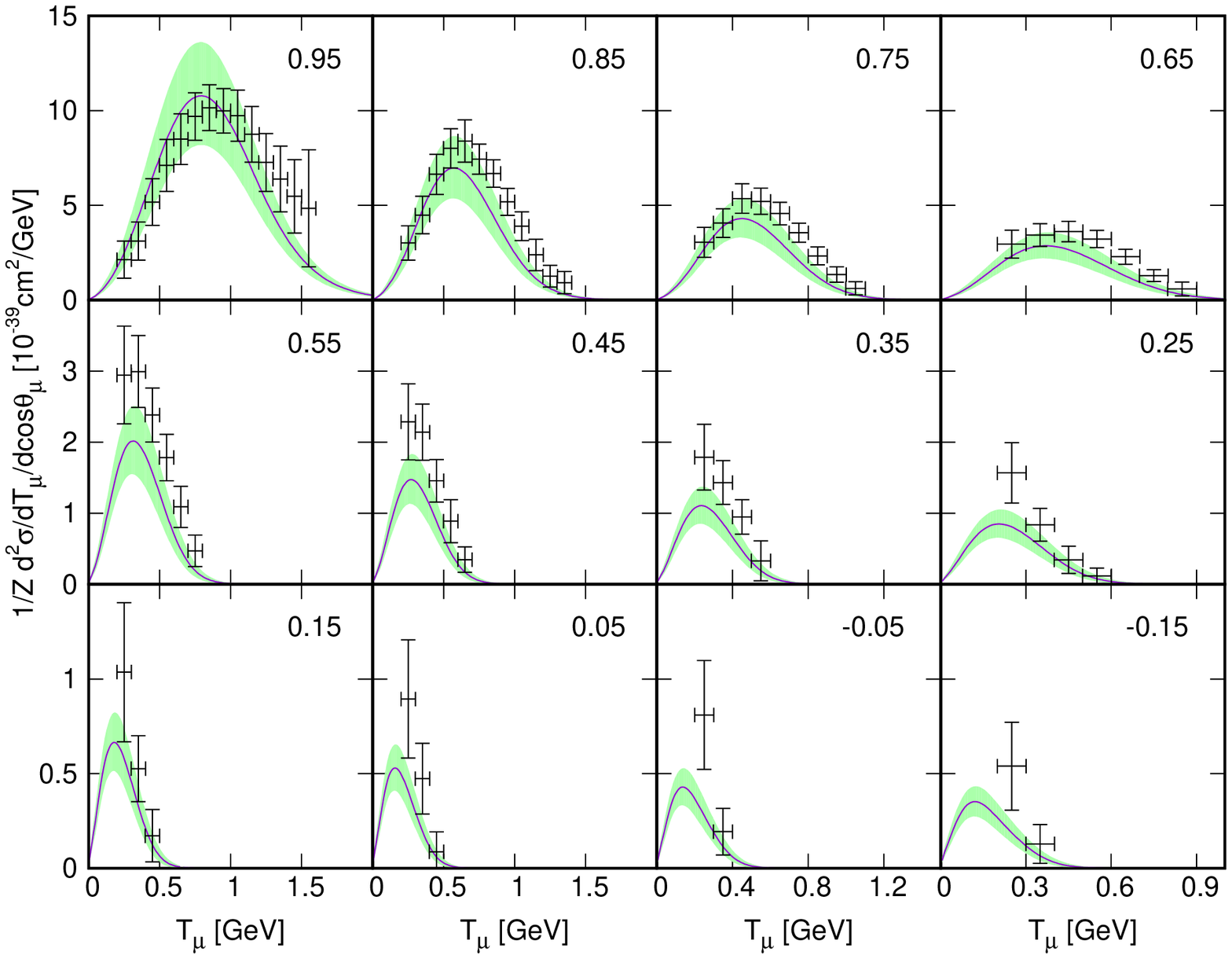}
\caption{ (Color online) Flux-integrated double-differential cross
  section per target proton for the CCQE $(\bar{\nu}_\mu,\mu^+)$
  reaction on $^{12}$C in the SuSAM* model. 
Each panel is labeled by
  the mean value of $\cos\theta_\mu$ in each experimental bin. 
The
  experimental data are from \cite{Agu13}. }\label{antinu-miniboone}
\end{figure*}

In this section we present our results for the quasielastic neutrino
and antineutrino cross sections on $^{12}$C and $^{16}$O within the
SuSAM* model.  The parameters of the model are the effective mass
$M^*=0.8$ and the Fermi momentum $k_F=225$, 230 MeV/c, for carbon and
oxygen, respectively. The other input of the model is the
phenomenological scaling function $f^*(\psi^*)$, extracted from
$(e,e')$ data as described in the previous section.

\subsection{MiniBooNE}

We start with the discussion of the MiniBooNE results \cite{Agu10,Agu13}.
In Fig. 2 we show our results for the flux-averaged double differential
 cross section
\begin{equation} \label{average}
\frac{d^2\sigma}{dT_\mu d\cos\theta_\mu}
=
\frac{1}{\Phi_{tot}}
\int dE_\nu \Phi(E_\nu)
\frac{d^2\sigma
}{dT_\mu d\cos\theta_\mu}(E_\nu) \, , 
\end{equation}
where 
$\frac{d^2\sigma
}{dT_\mu d\cos\theta_\mu}(E_\nu)$
is the computed cross section for fixed neutrino energy $E_\nu$.
The neutrino flux $\Phi(E_\nu)$ corresponds to the MiniBooNE experiment
on $^{12}$C nucleus \cite{Agu10}. 

In the figure the band predictions and the central cross section
values are compared to the experimental data of Ref \cite{Agu10},
which are given in bins of $\cos\theta_\mu$ and $T_\mu$. In each panel
of Fig. \ref{nu-miniboone} we fix the $\cos\theta_\mu$ to the
corresponding experimental bin and the cross section is plotted as a
function of $T_\mu$. In each panel we perform an integration over the
corresponding angular window in $\cos\theta_\mu$ and divide by the bin
width $\Delta\cos\theta_\mu=0.1$.

We observe that the strength and peak position are reasonably well
described by our model, and that the band thickness is similar to the
experimental errors. The data are slightly above our central curves,
except for very forward angles ($\cos\theta_\mu \sim 0.95$
panel). Thus, we find that, in general, the data are consistent with
the band within the experimental errors. Note that, by construction,
the band contains by definition the purely QE nuclear effects coming
from the $(e,e')$ data. This is a very rewarding result which
essentially confirms in a quantitative way the underlying hypothesis
of the scaling analysis, namely, the fact that around the QE the main
difererence between electron and neutrino scattering is {\it mainly}
due to the different currents and not so much on the intricacies of
nuclear effects. Of course, such a description has limitations and is
subjected to improvements. Actually, the experimental neutrino data
above the QE band indicate the existence of QE-like effects without
pions, as 2p-2h meson-exchange currents (MEC) and short-range
correlations or $\pi$-emission and reabsorption. Besides, the neutrino
data falling slightly below the band for low muon energy in the top
left panel point to low-$q$ mechanisms which are, in general,
overestimated by the scaling model. All these effects are known to
violate scaling and are expected to produce additional contributions
to the band results. Our goal here is limited to study the
implications of the $(e,e')$ band when directly translated to neutrino
scattering.  The inclusion of these effects, while extremely
interesting, is beyond the scope of this work and is left for future
research.

\begin{figure*}
\centering
\includegraphics[width=\textwidth]{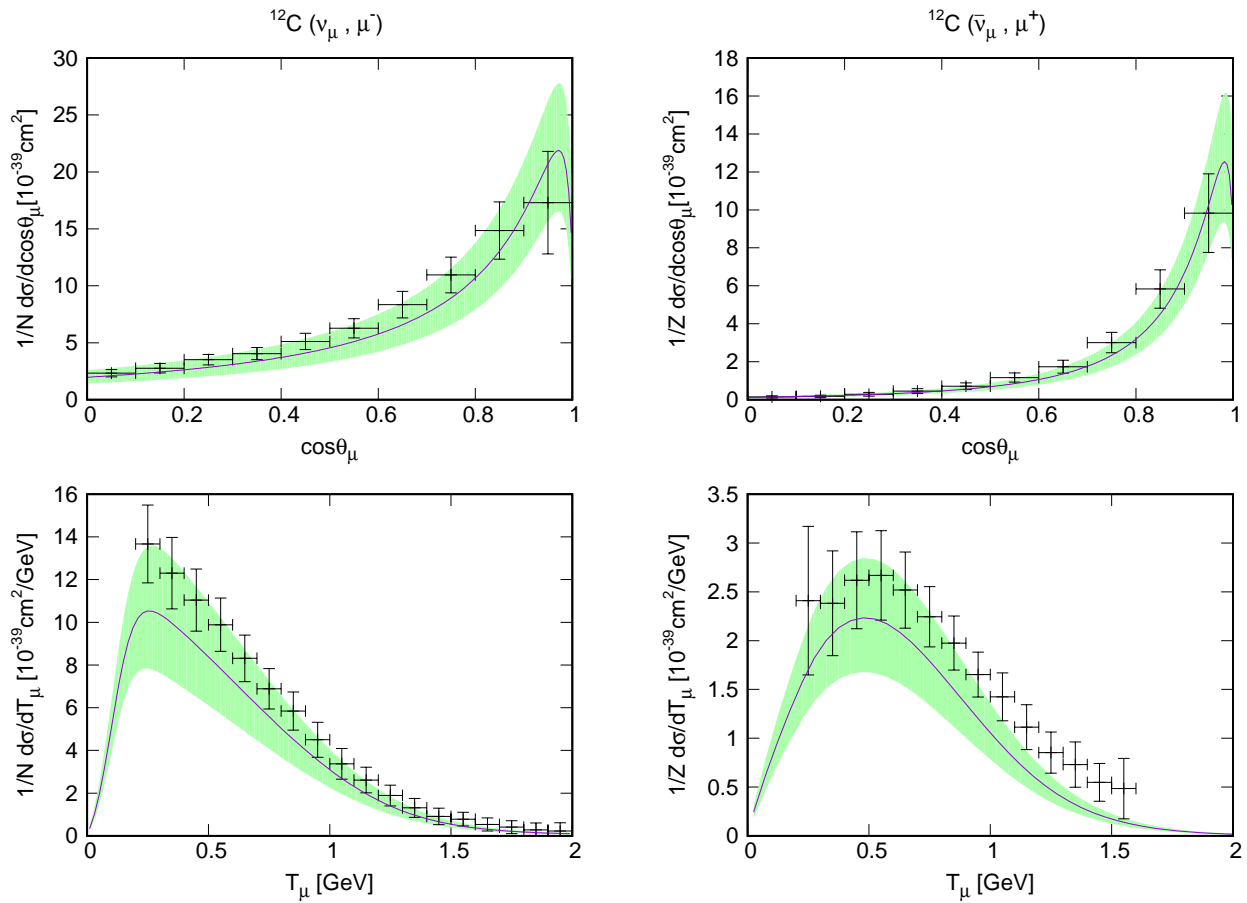}
\caption{ (Color online) Flux-integrated single-differential cross
  sections per target neutron (proton) for the CCQE neutrino
  (antineutrino) reactions on $^{12}$C in the SuSAM* model. Left
  panels are for neutrinos and right ones for antineutrinos. The
  experimental data are from \cite{Agu10,Agu13}
}\label{single-miniboone}
\end{figure*}

\begin{figure*}
\centering
\includegraphics[width=\textwidth]{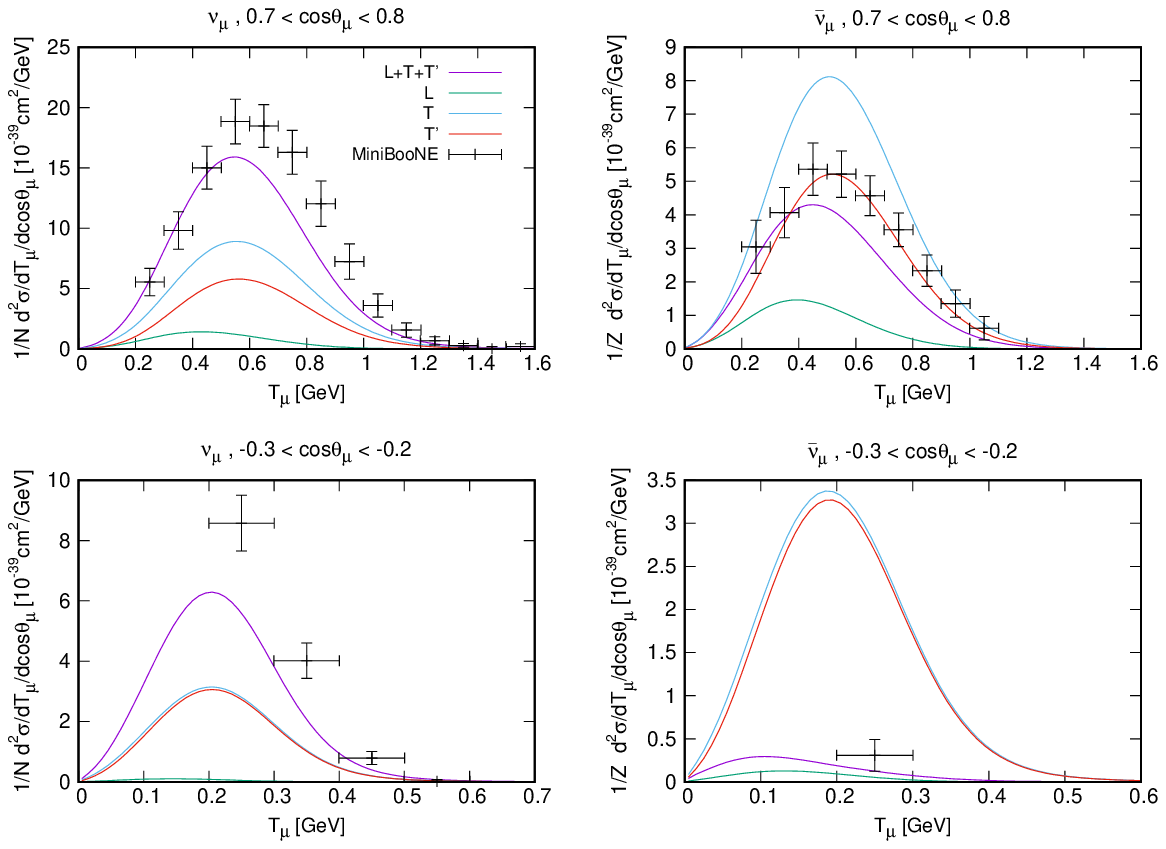}
\caption{ (Color online) Flux-integrated double-differential cross
  section per target neutron (proton) for the CCQE neutrino
  (antineutrino) reaction on $^{12}$C in the SuSAM* model (neutrino
  left and antineutrino right).  The separate contributions of the L, T
  and T' responses are highlighted for two selected kinematics.
 The experimental data are from
  \cite{Agu10,Agu13} }
\label{udias-miniboone}
\end{figure*}

The antineutrino double differential cross sections for the kinematics
of the MiniBooNE experiment \cite{Agu13} are shown in
Fig. \ref{antinu-miniboone}. Again a good description of data is
observed. In general the data are above our central curves, leaving
room for additional nuclear effects not included in the SuSAM* model.

Our results on Figs. \ref{nu-miniboone} and \ref{antinu-miniboone} are
quantitatively slightly larger than those of the RMF model of Udias
{\it et al.}  \cite{Ama11,Iva13}, but it is apparent that the RMF is
closely inside our uncertainty band. This was to be expected because
both models are based on the same theoretical Walecka model
\cite{Ser86}. The main difference between both approaches is that the
model of Udias {\it et al.}  describes finite nuclei with local vector
and scalar potentials, while here our potentials are constant and
generate a fixed effective mass inside the volume containing the
RFG. In our case the scaling function takes into account the finite
size of the nucleus and, besides, it is phenomenological, what
accounts for the differences between both models.

Despite the fact that our model is not including 2p-2h explicitly, it
is remarkable that the agreement of our central curve with the
MiniBooNE data is similar to that obtained with more sophisticated
models as those by authors Nieves {\em et al.}  \cite{Nie12,Nie13},
Martini {\em et al.}  \cite{Mar11,Mar13}, and Mosel {\em et al.}
\cite{Gal16}.  This is so because the RMF includes some dynamical
relativistic effects like enhancement of transverse response due to
lower components of nucleon spinors and other nuclear effects hidden
into the phenomenological scaling function $f^*(\psi^*)$.

The uncertainty imposed by the QE $(e,e')$ data 
over neutrino scattering can be
globally appreciated in the single differential cross sections of Fig.
\ref{single-miniboone}, obtained by integration of the double
differential ones. In general the data are within the SuSAM* band.
Our central curves are systematically slightly below the data, in
agreement with the results of the previous Figs. 2 and 3.  The thickness of
the band seems larger than in the $(e,e')$ cross
sections. This appears to be due to the integration over the neutrino
flux, which mixes the bands for different kinematics.

A closer insight into our results is considered in
Fig. \ref{udias-miniboone}. Here we show the separate contributions
of the $L$, $T$ and $T'$ responses to the cross section for several
kinematics. The contribution of $L$ response is considerably smaller
than the transverse channels. The $T$ response gives the largest
contribution. This indicates that the axial contribution to the
transverse cross section is larger than the $VV$ one.  For large
scattering angle the $T$ and $T'$ responses tend to be almost equal.
This produces a large cancellation in the antineutrino cross section,
which is therefore very small for large angles.  Due to this
cancellation one would expect the antineutrino cross section to be more
sensitive to the details of the longitudinal responses. The results of
Fig. \ref{udias-miniboone} are useful to further be compared with 
those of the RMF model of Udias {\it et al.} (see Figs. 2 and 3 of
ref. \cite{Iva13}), being in fair agreement with our findings.  This
comparison again ensures the similitudes between the SuSAM* and the RMF
in finite nuclei for intermediate energies.

\begin{figure*}
\centering
\includegraphics[width=\textwidth]{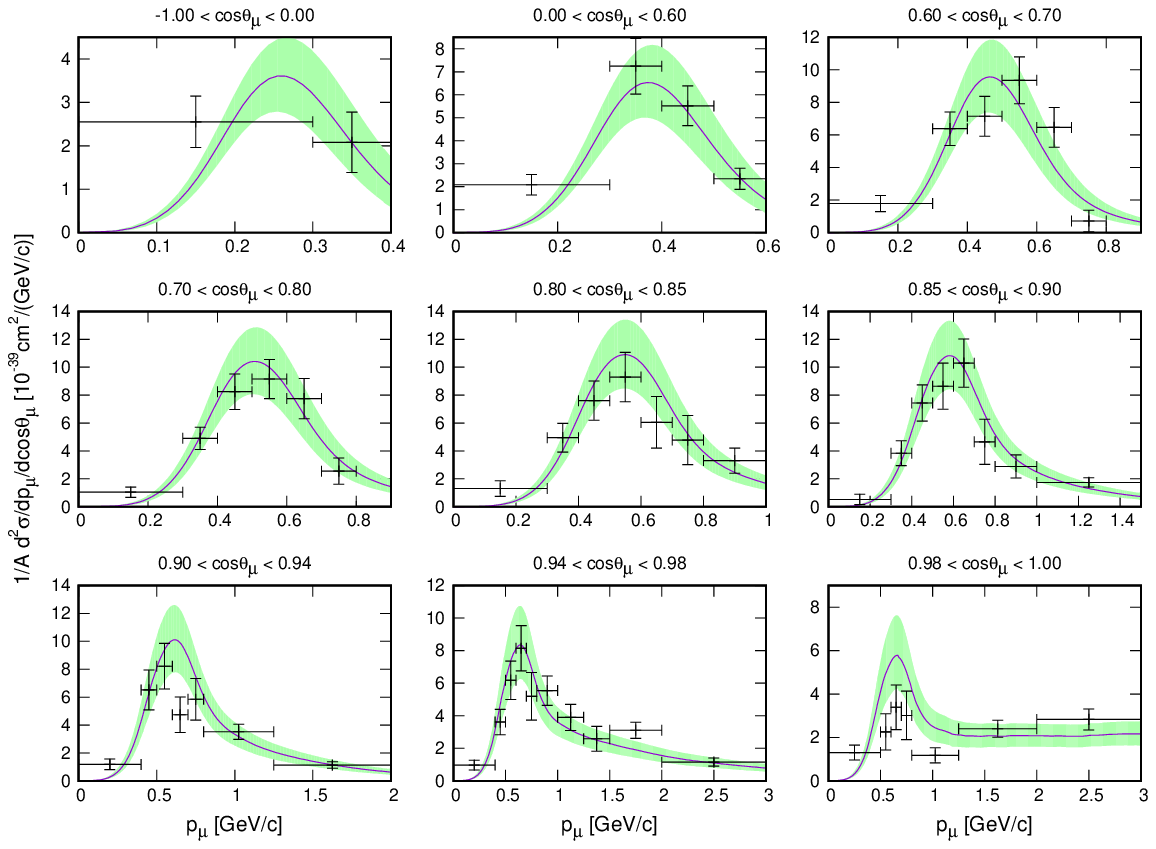}
\caption{T2K flux-folded double differential CCQE cross section per
  nucleon for $\nu_\mu$ scattering on $^{12}$C in the SuSAM*
  model. Experimental data are from \cite{Abe16}. }\label{T2K-qenu}
\end{figure*}

\begin{figure*}
\centering
\includegraphics[width=\textwidth]{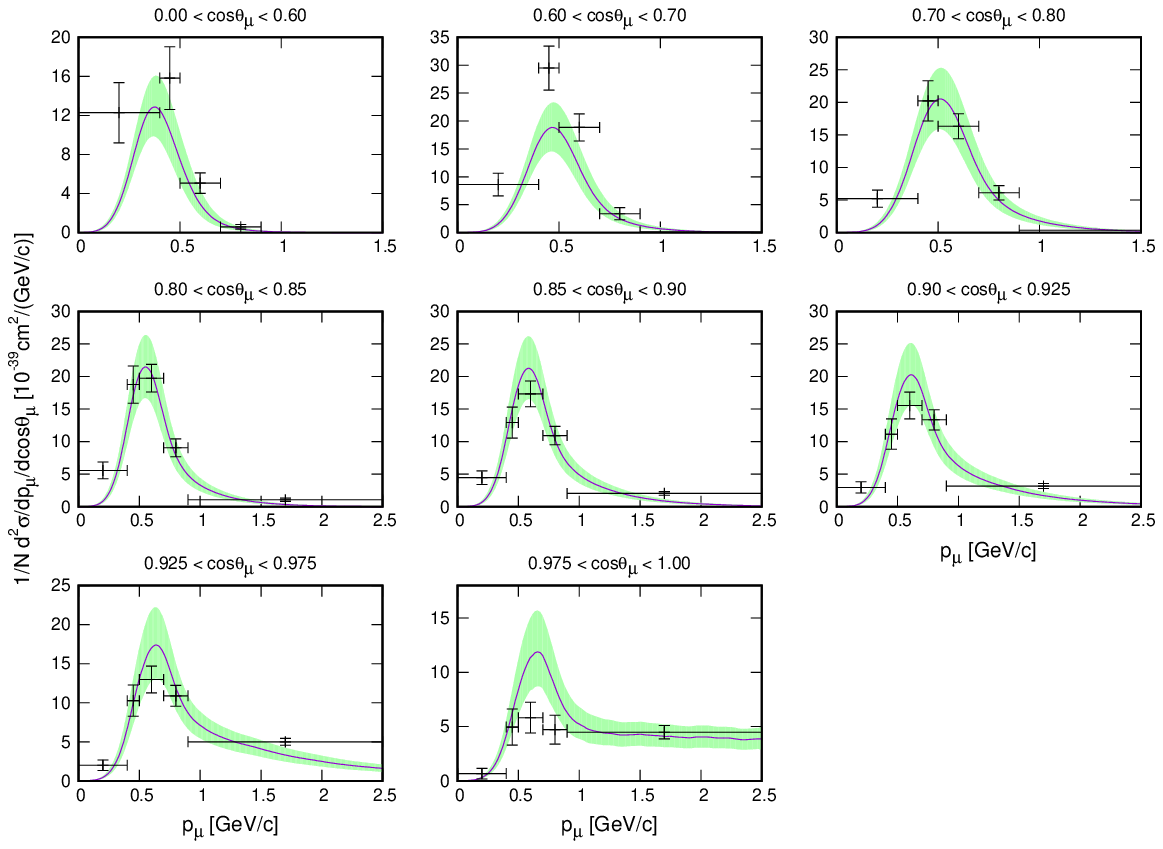}
\caption{T2K flux-folded double differential CCQE cross section per
  neutron for $\nu_\mu$ scattering on $^{16}$O in the SuSAM*
  model. Experimental data are from \cite{Abe17}. }\label{T2K-oxigeno}
\end{figure*}

\begin{figure}
\centering
\includegraphics[width=8cm]{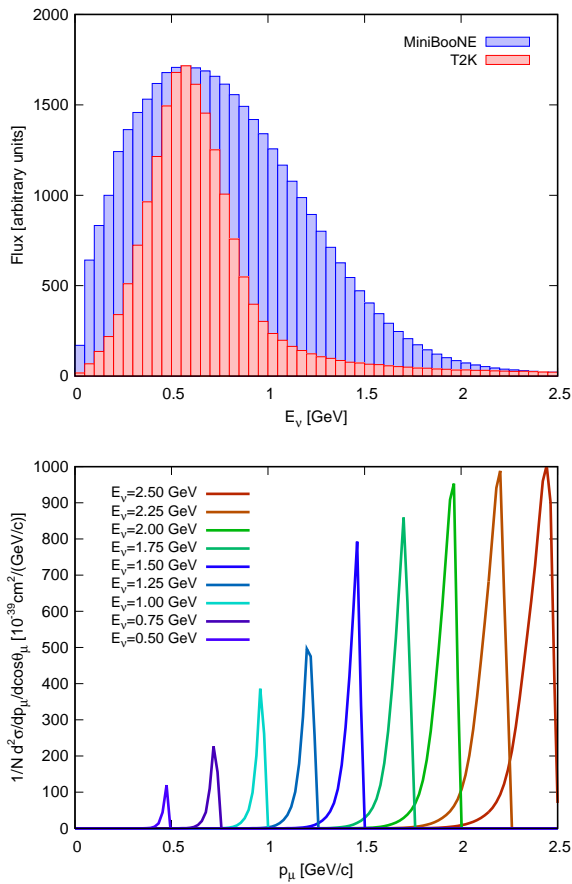}
\caption{(color online) 
 Top panel:  Comparison of MiniBooNE and T2K neutrino fluxes.
Bottom panel: 
 Double-differential  
 CCQE    $^{16}$O  $(\nu_\mu,\mu^-)$  cross section 
for fixed neutrino energies computed  for the angular bin 
$0.975 < \cos\theta_\mu < 1$. 
}
\label{efija}
\end{figure}

\begin{figure*}
\centering
\includegraphics[width=\textwidth]{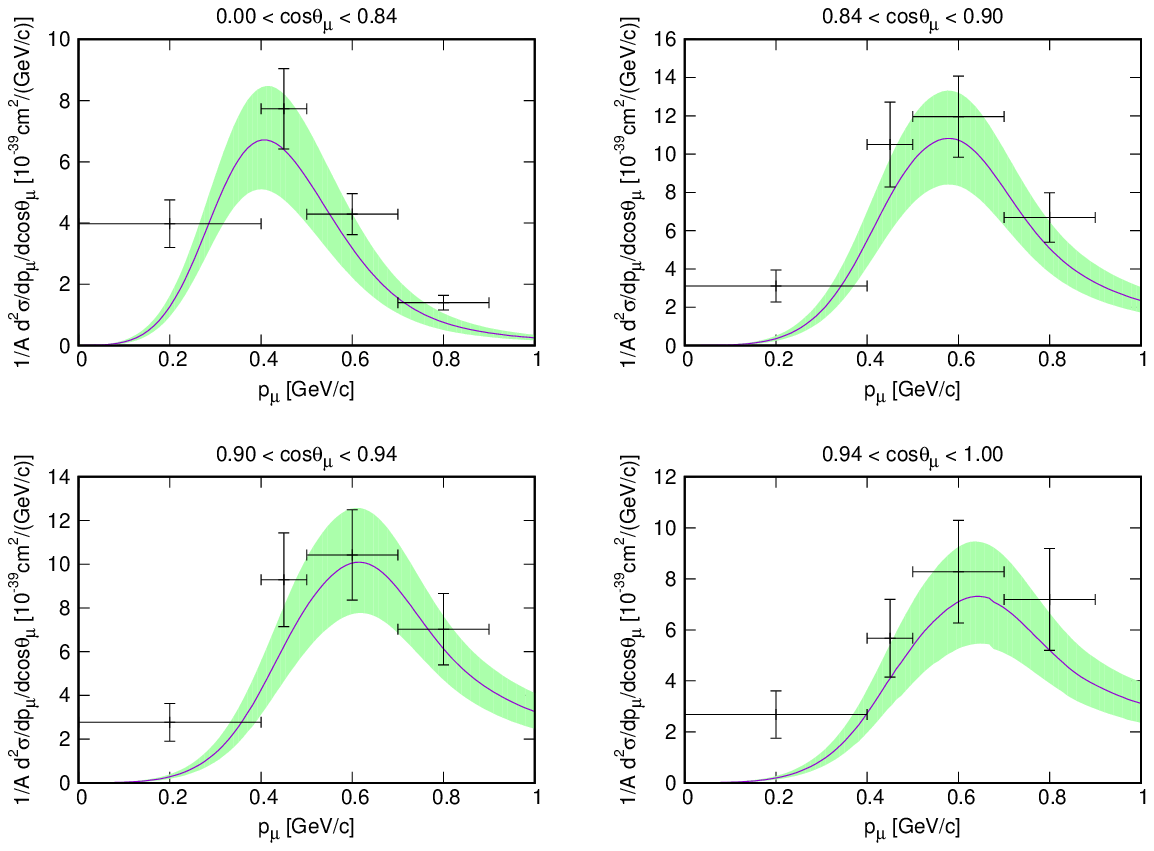}
\caption{T2K flux-folded double differential CC inclusive cross section per
  nucleon for $\nu_\mu$ scattering on $^{12}$C in the SuSAM*
  model. Experimental data are from \cite{Abe13}. }
\label{T2K-inclusive-nu}
\end{figure*}

\begin{figure}
\centering
\includegraphics[width=8cm]{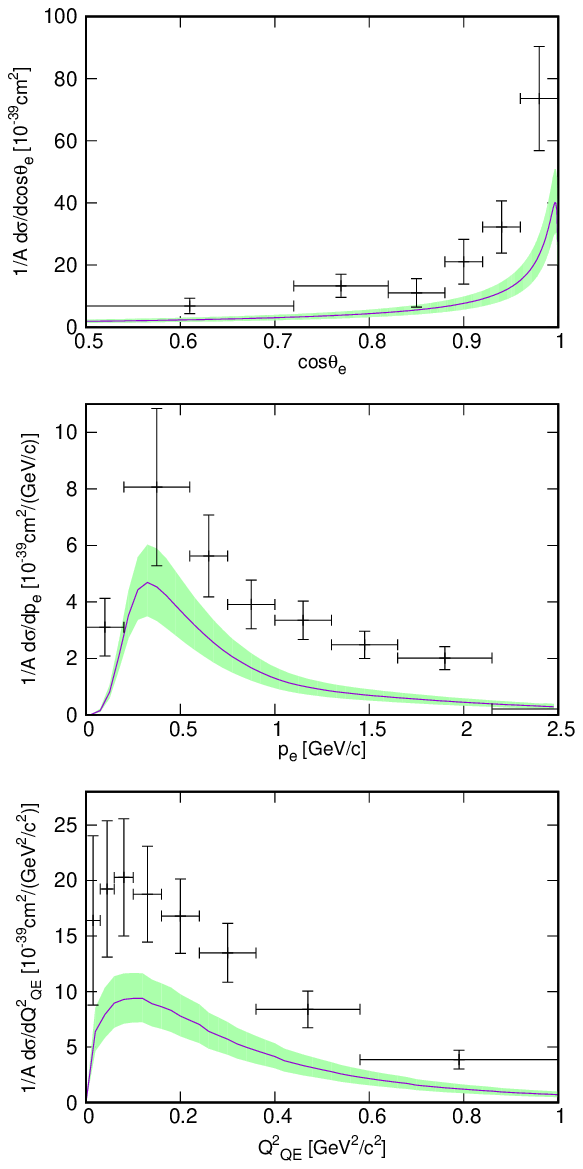}
\caption{T2K flux-folded single differential CC inclusive cross section per
  nucleon for $\nu_e$ scattering on $^{12}$C in the SuSAM*
  model. 
 Experimental data are from \cite{Abe14}.
The neutron binding energy for this case  is $E_B=25$ MeV.
 }
\label{T2K-nue}
\end{figure}

\subsection{T2K}

In Figs. \ref{T2K-qenu} and \ref{T2K-oxigeno} we compare the SuSAM*
predictions with the measurement of double-differential muon neutrino
CC cross section without pions (CCQE-like) of the T2K experiment from
$^{12}$C \cite{Abe16} and $^{16}$O \cite{Abe17}.  The experimental
data nicely fall inside the QE uncertainty band except for very
forward angles, where the data are overestimated around the maximum of
the cross section.  This is related to the limitations of the SuSAM*
model to describe the low momentum transfer region.

 In contrast to the MiniBooNE experiment, the T2K
data bins are not equally-spaced in $\cos\theta_\mu$, and the forward
angle dependence is probed in some more detail than in the MiniBooNE
analysis. Note that in all the calculations we average the cross
section over the corresponding angular bin. The neutrino energy
distribution of the T2K experiment is also different from that of the
MiniBooNE experiment, the former being narrower around the maximum
value $\sim 0.5$ GeV. Both neutrino fluxes are compared in
Fig. \ref{efija}.

The flux folding of the cross section, Eq. (\ref{average}),
implies an integration over the incident neutrino energy, which
produces a smearing of the cross section for different energies.
This results in a mild model dependence of the QE neutrino cross sections.
Being the flux narrower than the MiniBooNE one,  
one would expect the T2K experiment to be more suited
for discriminating over different theoretical models of the reaction.
Note also that the T2K angular bins are smaller for small angles
($\cos\theta_\mu > 0.8$), but for large angles ($\cos\theta_\mu <0.6$)
the bins are embracing a large angular sector, thus acquiring a
larger additional smearing of angular cross sections.  For all these reasons,
the T2K and MiniBooNE cross sections are not directly comparable
because they explore different energy and angular regions of the QE
peak by selected integrations.

In Fig. \ref{efija} we analyze in more detail the smearing effect
produced by the folding with the neutrino flux. We show the QE cross
section for fixed neutrino energies $E_\nu=0.5, \ldots, 2.5$ GeV, as a
function of the muon momentum for the kinematics $0.98 <
\cos\theta_\mu < 1$ (corresponding to the last panels in Figs.
\ref{T2K-qenu} and \ref{T2K-oxigeno}).  For fixed neutrino energy, the
cross section is a narrow peak contributing to the flux-folded cross
section only in a narrow region in $p_\mu$.  As $E_\nu$ increases the
allowed muon energy increases and the peak position moves to the
right, and its width increases, as a function of $p_\mu$. 
From this figure we can infer that the error of the 
reconstruction of the neutrino energy from the flux-folded 
cross section is related to the width of these peaks for 
fixed neutrino energy ---or more precisely 
for fixed $T_\mu$, which is of the same order.

 It is interesting to see that the strength of the cross section
 increases with $E_\nu$. This is in contrast to the electron
 scattering QE cross section, which decreases with the incident
 electron energy.  The reason is because in the electromagnetic
 interaction there is a photon propagator squared $1/Q^4$, which
 decreases with $Q^2$, while in the weak CC case there is a $W^{\pm}$
 propagator squared $\simeq 1/M_W^4$ which is almost constant for the
 low $Q^2$-values considered here.  Therefore the neutrino cross
 section increases with $E_\mu$.  This increase is partially balanced
 with a decrease of the neutrino flux inside the
 folding. Accidentally, this balance turns out to be almost perfect in
 the case of the T2K experiment for the kinematics of Fig. \ref{efija}
 and that is the reason why for large $p_\mu$ the flux-folded cross
 section does not fall to zero and is almost constant in the last
 panels of Figs. \ref{T2K-qenu} and \ref{T2K-oxigeno}.

To complete the comparison with T2K data, 
in Figs. \ref{T2K-inclusive-nu} and 
\ref{T2K-nue} we compare our model 
with the data of total inclusive cross sections for $\nu_\mu$
\cite{Abe13} and $\nu_e$ CC scattering from $^{12}$C \cite{Abe14}.
Up to now we have compared with inclusive data without pions in the
final state (CCQE-0$\pi$), corresponding to QE-like events, but they are
not restricted to one-nucleon emission because these semi-inclusive
cross sections contain multi-nucleon emission, mainly from 2p-2h final
states.  Our model accounts for this multi-nucleon emission, at least
partially, because the effective mass and scaling function is
extracted directly from cross section data. The width of the band
should account for those processes that violate scaling, but remain
close to the QE peak. In particular, the tail in the scaling function
is produced by nucleons with large momentum (compared to $k_F$) inside
the ground state, which are mainly produce by short-range
correlations. Other mechanisms such as meson-exchange currents also
should contribute partially to the uncertainty band.

In addition to this, the inclusive neutrino data in
Figs. \ref{T2K-inclusive-nu} and
\ref{T2K-nue} contain also explicit pion emission and other inelasticities, 
which do not scale and have been disregarded in our selection of QE
$(e,e')$ data. That is why our results underestimate the inclusive
cross sections, as it is more apparent in the case of $\nu_e$, shown
in Fig. \ref{T2K-nue}.  

\begin{figure}
\centering
\includegraphics[width=9cm]{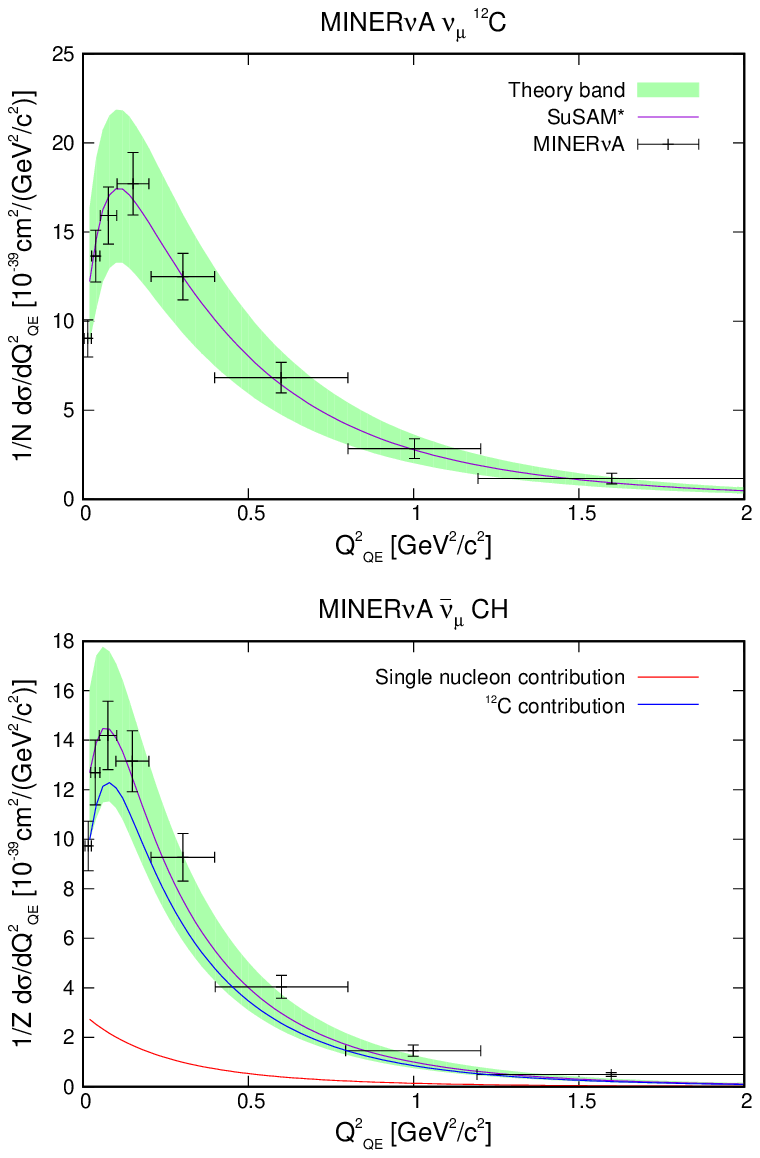}
\caption{(color online)
 Flux-folded CCQE $(\nu_\mu,\mu^-)$ and
  $(\bar{\nu}_\mu,\mu^+)$ scattering from $^{12}$C and CH scattering
  compared to the MINERvA experiment.  The H contribution is
  obtained from the elastic antineutrino-proton cross section divided
  by $Z=7$. The data are from \cite{Bet16}.}
\label{minerva}
\end{figure}

\begin{figure*}
\centering
\includegraphics[width=\textwidth]{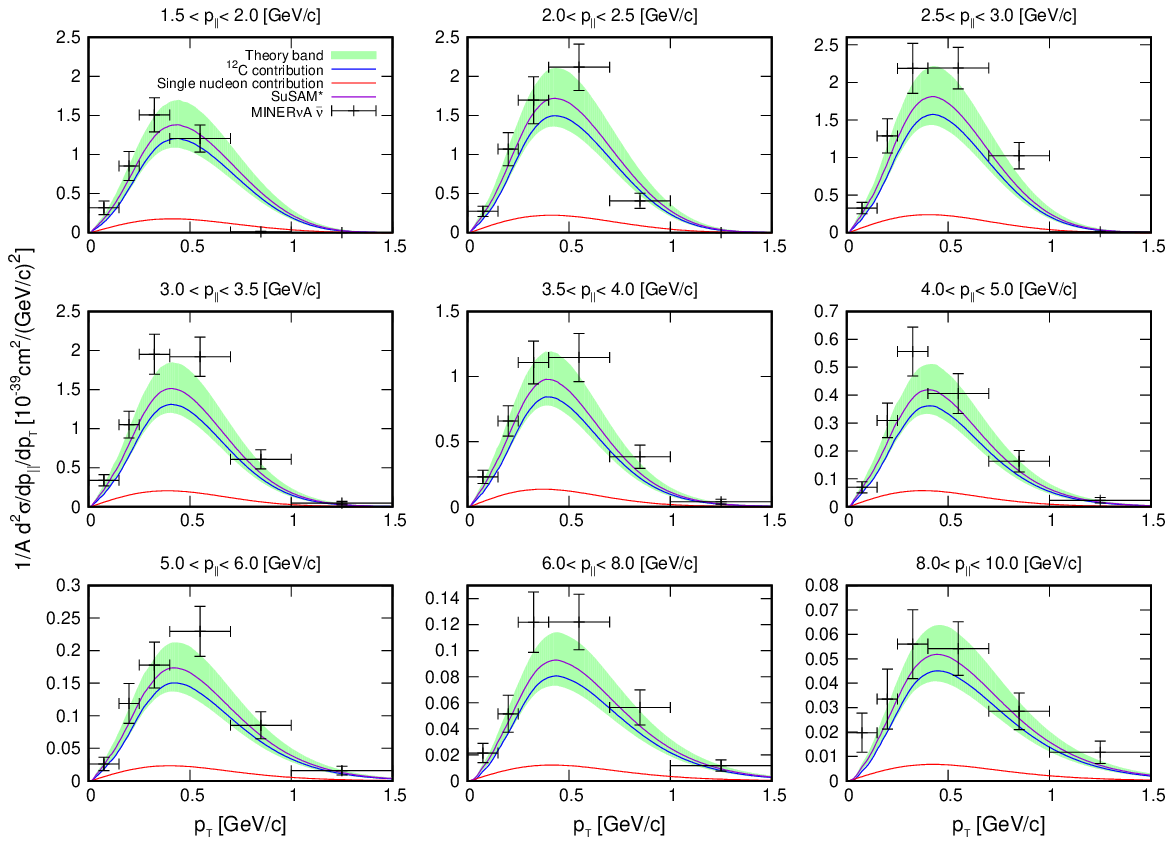}
\caption{(color online)
 Flux-folded 
double-differential cross section 
$\frac{d^2\sigma}{d p_{\parallel} dp_{\perp}}$
  for antineutrino CCQE  scattering from CH  (hydrocarbon)
  compared to the MINERvA experiment.  The H contribution is
  obtained from the elastic antineutrino-proton cross section divided
  by $A=13$. The data are from \cite{Pat18}.}
\label{minerva2}
\end{figure*}

\subsection{MINERvA}

In Fig. \ref{minerva} we show the
flux-folded CCQE single-differential cross section $d\sigma/dQ^2_{QE}$
for $(\nu_\mu,\mu^-)$ and $(\bar{\nu}_\mu,\mu^+)$ scattering from
$^{12}$C and CH respectively, compared to the MINERvA experiment
\cite{Bet16}.  This cross section is presented as a function of the
reconstructed variable $Q^2_{QE}$, which is not the true $Q^2$, but it
is computed from the reconstructed neutrino energy assuming
quasielastic scattering from a nucleon at rest.  In Appendix A we show
details on how this cross section is computed.  The cross section data
for antineutrino scattering contain the H contribution from the
target. In Appendix B we show how we evaluate this cross section
within our formalism.  As we can see in Fig. \ref{minerva}, all the
MINERvA data fall with our uncertainty band and, moreover, they are
very close to the central value of the SuSAM* model predictions.

To finish our discussion, in Fig. \ref{minerva2} we show the double
differential antineutrino cross section for QE scattering on CH
(hydrocarbon) corresponding to the recent measurements in the MINERvA
detector \cite{Pat18}. 

The double-differential measurements of antineutrino QE
scattering in the MINERvA detector provide a complete description of
observed muon kinematics with respect to the muon longitudinal and
transverse momentum  with respect to the incident neutrino.
These are related to the 
variables $p_\mu, \cos\theta_\mu$ by
\begin{eqnarray}
p_{\parallel} &=& p_\mu\cos\theta_\mu \\
p_{\perp} &=& p_\mu\sin\theta_\mu 
\end{eqnarray}
It is then straightforward to compute the Jacobian of the transformation
to the usual cross section variables  $E_\mu,\cos\theta_\mu$ 
\begin{equation}
d p_{\parallel} dp_{\perp}= \frac{E_\mu}{\sin\theta_\mu}
d E_\mu d\cos\theta_\mu
\end{equation}
Therefore
 we compute the MINERvA double-differential cross section as
\begin{equation}
\frac{d^2\sigma}{d p_{\parallel} dp_{\perp}} =
\frac{\sin\theta_\mu}{E_\mu}
\frac{d^2\sigma}{d E_\mu d\cos\theta_\mu}
\end{equation}

Our calculation in Fig. \ref{minerva2} is compared with the 'true'
CCQE data from tables XXII, XXIII and XXIV of ref. \cite{Pat18}.  Most of the
data are within the uncertainty band and slightly above the central
SuSAM* results. This may indicate the presence of some non-QE events
in the data.  The hydrogen contribution is also included in the
calculation from hydrocarbon. It is found to provide a small
correction which slightly increases the cross section, but it could be
safely ignored without sensibly modifying the uncertainty band of
Fig. \ref{minerva2}. Note that in each panel of the figure the
parallel component of the muon momentum is averaged over the
corresponding experimental bin as indicated. 
By plotting the cross section 
as a function of  the perpendicular muon momentum not only 
the total muon momentum increases, but also the 
angle, opposite to the usual
$E_\mu$-plots where the $\cos\theta_\mu$ bins are kept constant.

\section{Conclusions}

Summarizing, in this work we have provided predictions for the CCQE
neutrino and antineutrino cross sections and their theoretical
uncertainties. These uncertainties are extracted directly from the
$(e,e')$ data and can be considered upper limits to the expected
systematic errors coming from the nuclear modeling of the reaction.
We have compared our model with the available QE neutrino and
antineutrino double-differential and differential cross section data
from MiniBooNE, T2K and MINERvA experiments, with a reasonable
agreement.  The theoretical uncertainty bands are around $20\%-30\%$
and, in general, of the same order as the experimental errors.

The agreement of our central results with data is similar to that
obtained with much more sophisticated theoretical models.  We emphasize
that the results were obtained without any tuning of model
parameters except the relativistic effective mass and the Fermi
momentum, extracted from a large body of $(e,e')$ data around the QE.
The accord with data over different experiments,
different nuclei, and spanning a wide range of neutrino and
antineutrino energies, shows that the SuSAM* uncertainty bands
faithfully encode intricate nuclear effects and qualify as a suitable
tool for the validation of models of QE-like interactions.  We plan to
explore ways to reduce the systematic errors, for instance by
combining the SuSAM* model with explicit additional contributions from
meson-exchange currents, which only are partially included in the
phenomenological scaling function, or by extending the model to
account for the inelastic channels.

\section{Acknowledgements}

This work has been partially supported by the Spanish Ministerio de
Economia y Competitividad (grants Nos. FIS2014-59386-P and
FIS2017-85053-C2-1-P) and by the Junta de Andalucia (grant
No. FQM-225). V.L.M.C.  acknowledges a contract with Universidad de
Granada funded by Junta de Andalucia and Fondo Social Europeo.

\appendix

\section{Reconstructed $Q^2_{QE}$ differential cross section}

A quantity of interest in the experiments is the reconstructed
neutrino energy assuming quasielastic scattering on a nucleon at rest.
By energy-momentum conservation, it is given by \cite{Agu10,Pat18, 
Pat18_thesis}
\begin{equation} \label{enuqe}
E_{\nu}^{QE}=
\frac{m_p^2-m'_n{}^2-m_\mu^2+2m'_nE_\mu}
{2(m'_n-E_\mu+p_\mu\cos\theta_\mu)}
\end{equation}
where $m_n'=m_n-E_b$, and $E_b=34$ MeV is an effective binding energy for an initial neutron
at rest in $^{12}$C.  For the inverse reaction case of antineutrinos
one should exchange the neutron and proton masses in the formula and
a change in $E_b=30$ MeV.
With this, the reconstructed $Q^2_{QE}$ is given by 
\begin{equation}\label{q2qe}
Q^2_{QE}=2E_{\nu}^{QE}(E_\mu-p_\mu\cos\theta_\mu)-m_\mu^2
\end{equation}
Note that the reconstructed variable $Q^2_{QE}$ is not the true $Q^2$
because the true neutrino energy is unknown. It is a function of
$(\cos\theta_\mu,E_\mu)$. It is a convenient variable to present results in a
representation similar to the neutrino-nucleon cross section.  The
definition of the cross section $d\sigma/dQ^2_{QE}$ is the following
\begin{equation}
\frac{d\sigma}{dQ^2_{QE}}
=
\int dE_\mu 
\frac{ \displaystyle \frac{d\sigma}{d\cos\theta_\mu dE_\mu}  }
   {\left| \displaystyle \frac{\partial Q^2_{QE}}{\partial \cos\theta_\mu}\right| }
\end{equation}
were the cross section in the numerator, is the flux-folded
double-differential cross section. The denominator is the Jacobian in
the change of variables $(\cos\theta_\mu,E_\mu) \rightarrow
(Q^2_{QE},E_\mu)$. 
Therefore inside the integral the $\cos\theta_\mu$
variable is computed from Eqs. (\ref{enuqe}--\ref{q2qe}) by solving
for $\cos\theta_\mu$ in terms of the independent variables $(Q^2_{QE},E_\mu)$:
\begin{eqnarray}
\cos\theta_\mu
&=& 
\left[ E_\mu(m_p^2-m'_n\mbox{}^2-m_\mu^2+2m'_nE_\mu)  \right.
\nonumber\\
&&
- \left. (m_n'-E_\mu)(Q^2_{QE}+m_\mu^2) \right]
\nonumber\\
&&
\times
\left[ p_\mu(Q^2_{QE}+m_p^2-m'_n\mbox{}^2+2m'_nE_\mu)\right]^{-1}
\end{eqnarray}
The Jacobian in the denominator 
is then given by
\begin{eqnarray}
\kern -3mm 
\frac{\partial Q^2_{QE}}{\partial \cos\theta_\mu} 
&=& 
2 \frac{\partial E_\nu^{QE}}{\partial\cos\theta_\mu}(E_\mu-p_\mu\cos\theta_\mu)
-2 p_\mu E_\nu^{QE}
\\
\kern -3mm 
\frac{\partial E_\nu^{QE}}{\partial\cos\theta_\mu}
&=&
-\frac{m_p^2-m'_n\mbox{}^2-m_\mu^2+2m'_nE_\mu}
{2(m'_n-E_\mu+p_\mu\cos\theta_\mu)^2}
p_\mu.
\end{eqnarray}
The resulting differential cross section has been computed using (A3)
when we compare with the MINERvA data.

\section{The elastic $p(\bar{\nu}_\mu,\mu^+)n $ cross section}

Some experiments provide cross sections including the contributions
from the Hydrogen target atoms.  Therefore, when comparing with such
experiments, such as the MINERvA data, the individual proton cross
sections have to be added incoherently to the $^{12}$C one in
proportion to the relative number of protons in each species of the
target.

Here we compute this contribution as a particular case of the RFG
formalism for $k_F=0$, $Z=1$ and $M^*=1$. For the flux averaged antineutrino
cross section we have    
\begin{eqnarray}
\frac{d^2\sigma}{dT_\mu d\cos\theta_\mu}
&=&
\int dE_\nu \phi(E_\nu) \frac{m_n}{E'} \delta(E'+E_\mu-m_p-E_\nu)
\nonumber\\
&&
\times
\sigma_{0}(E_\nu)\sum_K V_K U^K(E',0),
\end{eqnarray}
where the elastic single nucleon responses $U^K(E',0)$ are evaluated
for a nucleon at rest using
Eqs. (\ref{ucc}, \ref{ucl}, \ref{ull}, \ref{ut}, \ref{utp}), for $k_F=0$,
or equivalently, by simply setting $\Delta=\tilde{\Delta}=0$ in those
equations. Being a nucleon at rest, the values of $q$ and $\omega$ are
calculated from the kinematics inside the integral. To
integrate over the neutrino energy using the Dirac delta,
we take into account that the final nucleon energy, $E'$, 
depends on the antineutrino energy through 
\begin{equation}
E'{}^2= m_n^2+E_{\nu}^2+p_\mu^2-2E_\nu p_\mu\cos\theta_\mu
\end{equation}
Differentiating we can write
\begin{equation}
\frac{dE_\nu}{E'}= \frac{d(E'-E_\nu)}{E_\nu-p_\mu\cos\theta_\mu-E'}.
\end{equation}
Now it is straightforward to integrate the Dirac delta, obtaining
\begin{eqnarray}
\frac{d^2\sigma}{dT_\mu d\cos\theta_\mu}
&=&
\phi(E_\nu) \frac{m_n}{|E_\mu-m_p-p_\mu\cos\theta_\mu|} 
\nonumber\\
&&
\times
\sigma_{0}(E_\nu)\sum_K V_K U^K(E',0),
\end{eqnarray}
where we have taken the absolute value of the Jacobian and 
replaced in the denominator $E_\nu-E'= E_\mu-m_p$.
The value of the antineutrino energy is
\begin{equation} 
E_{\nu}=
\frac{m_n^2-m_p^2-m_\mu^2+2m_pE_\mu}
{2(m_p-E_\mu+p_\mu\cos\theta_\mu)}
\end{equation}
and $E'= E_\nu-E_\mu+m_p$.
This solves the double-differential cross section problem.
Finally, to compute the single differential cross section 
$d\sigma/dQ^2_{QE}$ for the proton 
we apply again the method explained in Appendix A, for $E_b=0$.


\begin{thebibliography}{99}


\bibitem{Mos16} U. Mosel, Ann. Rev. Nuc. Part. Sci. 66 (2016), 171.

\bibitem{Kat17}
  T.~Katori and M.~Martini,
  J.\ Phys.\ G {\bf 45} (2018) no.1,  013001.

\bibitem{Pat18}
  C.~E.~Patrick {\it et al.} [MINERvA Collaboration],
  Phys.\ Rev.\ D {\bf 97} (2018) no.5,  052002.

\bibitem{Abe11} K. Abe et al. (T2K), Nucl. Instrm. Meth. {\bf A 659},
  106 (2011)

\bibitem{Abe18a} K. Abe et al. (T2K Collaboration), arXiv:1801.05148 [hep-ex].
\bibitem{Abe18b} K. Abe et al. (T2K Collaboration), arXiv:1802.05078 [hep-ex].

\bibitem{Ada16} P. Adamson et al. (NOvA Collaboration),
  Phys. Rev. Lett. 116, 151806 (2016).

\bibitem{Acc15} R. Acciarri et al. (DUNE Collaboration),
  arXiv:1512.06148 [physics.ins-det]

\bibitem{Pal16} 
  O.~Palamara [ArgoNeuT Collaboration],
  JPS Conf.\ Proc.\  {\bf 12}, 010017 (2016).

\bibitem{Acc14} 
  R.~Acciarri {\it et al.} [ArgoNeuT Collaboration],
  Phys.\ Rev.\ D {\bf 90}, no. 1, 012008 (2014)

\bibitem{Bac17} 
  A.~R.~Back {\it et al.} [ANNIE Collaboration],
  arXiv:1707.08222 [physics.ins-det].


\bibitem{Mar09} M. Martini, M. Ericson, G. Chanfray, J. Marteau, 
                 Phys.Rev. C80 (2009) 065501.

\bibitem{Nie11} J. Nieves, I. Ruiz Simo, M.J. Vicente Vacas, 
Phys.Rev. C83 (2011) 045501.

\bibitem{Gal16} 
  K.~Gallmeister, U.~Mosel and J.~Weil,
  Phys.\ Rev.\ C {\bf 94}, no. 3, 035502 (2016).

\bibitem{Meg16}
 G.D Megias, 
J.~E.~Amaro, M.~B.~Barbaro, J.~A.~Caballero, T.~W.~Donnelly and I. Ruiz Simo,
 Phys. Rev. D 94 (2016), 093004. 

\bibitem{Ama15}
  J.~E.~Amaro, E.~Ruiz Arriola and I.~Ruiz Simo,
  Phys.\ Rev.\ C {\bf 92} (2015) no.5,  054607
  doi:10.1103/PhysRevC.92.054607
  [arXiv:1505.05415 [nucl-th]].


\bibitem{Ama17} 
  J.~E.~Amaro, E.~Ruiz Arriola and I.~Ruiz Simo,
  Phys.\ Rev.\ D {\bf 95}, no. 7, 076009 (2017).



\bibitem{Alb88} W.M. Alberico, A. Molinari, T.W. Donnelly, E. L. Kronenberg, 
and J.W. Van Orden, Phys Rev. C 38 (1988) 1801.

\bibitem{Day90} 
  D.~B.~Day, J.~S.~McCarthy, T.~W.~Donnelly and I.~Sick,
  Ann.\ Rev.\ Nucl.\ Part.\ Sci.\  {\bf 40}, 357 (1990).


\bibitem{Don99} 
  T.~W.~Donnelly and I.~Sick,
  Phys.\ Rev.\ C {\bf 60}, 065502 (1999).

\bibitem{Ama05a} J.E. Amaro, M.B. Barbaro, J.A. Caballero, T.W. Donnelly
                 A. Molinari, and I. Sick,
                Phys. Rev. C {\bf 71}, 015501 (2005)

\bibitem{Ama05b} J.E. Amaro, M.B. Barbaro, J.A. Caballero, T.W. Donnelly,
                 C. Maieron,                  
                Phys. Rev. C {\bf 71}, 065501 (2005)

\bibitem{Meg16a} 
  G.~D.~Megias, J.~E.~Amaro, M.~B.~Barbaro, J.~A.~Caballero and T.~W.~Donnelly,
  Phys.\ Rev.\ D {\bf 94}, 013012 (2016).

\bibitem{Meg18} 
  G.~D.~Megias, M.~B.~Barbaro, J.~A.~Caballero, J.~E.~Amaro, T.~W.~Donnelly, I.~Ruiz Simo and J.~W.~Van Orden,
  arXiv:1711.00771 [nucl-th].

\bibitem{Ser86} B.D. Serot, and J.D. Walecka, Adv. Nucl. Phys. 16 (1986) 1.

\bibitem{Ros80} R. Rosenfelder, Ann. Phys. (N.Y:) 128, 188 (1980)



\bibitem{Ama17b} 
  V.L. Martinez-Consentino,  I.~Ruiz Simo,
J.~E.~Amaro, and E.~Ruiz Arriola 
  Phys.\ Rev.\ C {\bf 96},  064612 (2017).


\bibitem{Con18} 
  V.L. Martinez-Consentino,  I.~Ruiz Simo,
J.~E.~Amaro, and E.~Ruiz Arriola, 
  in preparation.





\bibitem{DeF83} 
  T.~De Forest,
  Nucl.\ Phys.\ A {\bf 392}, 232 (1983).


\bibitem{Mai02} 
  C.~Maieron, T.~W.~Donnelly and I.~Sick,
  Phys.\ Rev.\ C {\bf 65}, 025502 (2002)
  doi:10.1103/PhysRevC.65.025502
  [nucl-th/0109032].



\bibitem{Ang96} M. Anghinolfi {\em et al.,} Nucl. Phys. A 602, 405 (1996).

\bibitem{Con87} J.S. O'Connell {\em et al.,} Phys. Rev. C 35, 1063 (1987).



\bibitem{Agu10}
A.A. Aguilar-Arevalo, et al. (MiniBooNE Collaboration),  
Phys.Rev. D81 (2010) 092005. 

\bibitem{Agu13}
A.A. Aguilar-Arevalo, et al. (MiniBooNE Collaboration),  
Phys.Rev. D88 (2013) 032001. 


\bibitem{Iva13} M.V. Ivanov, R. Gonzalez-Jimenez, J.A. Caballero,
  M.B. Barbaro, T.W. Donnelly, and J.M. Udias, Phys. Lett. B 727
  (2013) 265.
  
  \bibitem{Ama11} 
J.E. Amaro, M.B. Barbaro, J.A. Caballero, T.W. Donnelly, and J.M. Udias, Phys. Rev. D 84, 033004 (2011).

\bibitem{Abe16} K. Abe et al., (T2K Collaboration),
Phys. Rev. D93 (2016) 112012.

\bibitem{Abe17} 
  K.~Abe {\it et al.} [T2K Collaboration],
  Phys.\ Rev.\ D {\bf 97}, no. 1, 012001 (2018).

\bibitem{Abe13} K. Abe et al. (T2K Collaboration), 
                Phys.Rev. D87 (2013) 9, 092003. 

\bibitem{Abe14} K. Abe et al. (T2K Collaboration), 
 Phys.Rev. Lett. 113, 241803  (2014).

\bibitem{Nie12} J. Nieves, I. Ruiz Simo, M.J. Vicente Vacas, 
Phys.Lett. B707 (2012) 72.


\bibitem{Nie13} J. Nieves, I. Ruiz Simo, M.J. Vicente Vacas, 
Phys.Lett. B721 (2013) 90.


\bibitem{Mar11} 
  M.~Martini, M.~Ericson and G.~Chanfray,
  Phys.\ Rev.\ C {\bf 84}, 055502 (2011)



\bibitem{Mar13} 
  M.~Martini and M.~Ericson,
  Phys.\ Rev.\ C {\bf 87}, no. 6, 065501 (2013)

\bibitem{Bet16} 
  M.~Betancourt,
  JPS Conf.\ Proc.\  {\bf 12}, 010016 (2016).
  doi:10.7566/JPSCP.12.010016

\bibitem{Pat18_thesis} C.E. Patrick, 
{\em 
Measurement of the 
Antineutrino Double-Differential Charged-Current Quasi-Elastic Scattering Cross Section at MINERvA}
Springer Theses, Springer Int. Publ. AG, 2018.

\end{thebibliography}
\end{document}